\documentclass{aa}  

\usepackage{graphicx}
\usepackage{txfonts}
\begin{document}

\title{Discovery of an activity cycle in the solar-analog HD 45184\thanks{Based on
data products from observations made with ESO Telescopes at the La Silla Paranal Observatory under
programs 072.C-0488 and 183.C-0972.}}

  \subtitle{Exploring Balmer and metallic lines as activity proxy candidates}

   \author{M. Flores
   \inst{1,5}
   \and          
    J. F. Gonz\'alez 
    \inst{1,3,5}
    \and
    M. Jaque Arancibia 
    \inst{1,5}          
    \and
    A. Buccino 
    \inst{2,4,5}
    \and
    C. Saffe 
    \inst{1,3,5}           
           }
    \institute{Instituto de Ciencias Astron\'omicas, de la Tierra y del Espacio (ICATE), Espa\~na Sur 1512,
    CC 49, 5400 San Juan, Argentina.
    \email{mflores,fgonzalez,mjaque,csaffe@icate-conicet.gob.ar}
    \and 
    Instituto de Astronom\'ia y F\'isica del Espacio (IAFE), Buenos Aires, Argentina.                    
    \email{abuccino@iafe.uba.ar}
    \and
    Facultad de Ciencias Exactas, F\'isicas y Naturales, Universidad Nacional de San Juan, San Juan, Argentina.
    \and
    Departamento de F\'isica, Facultad de Ciencias Exactas y Naturales, Universidad de Buenos Aires,
    Buenos Aires, Argentina.
    \and
    Consejo Nacional de Investigaciones Cient\'ificas y T\'ecnicas (CONICET), Argentina.
}

   \date{Received September 15, 1996; accepted March 16, 1997}

 
\abstract
{Most stellar activity cycles similar to that found in the Sun have been detected by using
the chromospheric \ion{Ca}{ii} H\&K lines as stellar activity proxies.
However, it is unclear if such activity cycles could be identified using other
optical lines.}   
{To detect activity cycles in solar-analog stars and determine if these can be identified
through other optical lines, such as Fe II and Balmer lines. We study the solar-analog star HD 45184
using HARPS spectra, whose temporal coverage and the high quality of the spectra
allow us to detect both long and short-term
activity variations.}
{We analyse the activity signatures of HD~45184 by using 291 HARPS spectra obtained between 2003 and 2014. 
In order to search for line-core fluxes variations, we focus on \ion{Ca}{ii} H\&K and Balmer H$\alpha$,
H$\beta$ lines, which are usually used as optical chromospheric activity indicators. We calculate the
HARPS-S index from \ion{Ca}{ii} H\&K lines and convert it to the Mount-Wilson scale. 
In addition, we also consider as activity indicators the equivalent widths of Balmer lines. 
Moreover, we analyse the possible variability of \ion{Fe}{ii} and other metallic lines in the optical spectra. The spectral variations are analysed for periodicity using the Lomb-Scargle periodogram.}
{We report for the first time a long-term 5.14-yr activity cycle in the solar-analog star HD~45184 derived from Mount Wilson S index. This makes HD 45184 one of most similar stars to the Sun with known activity cycle. Such variation is also evident in the first lines of the Balmer series, which not always show a correlation with activity in solar-type stars.
Notably, unlike the solar case, we also found that the equivalent widths of the
high photospheric \ion{Fe}{ii} lines (4924 \AA, 5018 {\AA} and 5169 \AA) are modulated ($\pm$ 2 m\AA) by the chromospheric cycle of the star.
These metallic lines show variations above 4$\sigma$ in the RMS spectrum, while some \ion{Ba}{ii} and \ion{Ti}{ii} lines
present variations at 3$\sigma$ level which could be considered as marginal variations.         
From short-term modulation of the S index we calculate a rotational period  of 19.98 days, which agrees
with its mean chromospheric activity level.
Then, we clearly show that the activity cycles of HD 45184 could be detected in both \ion{Fe}{ii} and Balmer
lines. 
}
{}


   \maketitle
%

\section{Introduction}

The pioneer research on stellar activity by \citet{1978ApJ...226..379W} and subsequent works
\citep[e.g.][]{1978PASP...90..267V,1991ApJS...76..383D,1995ApJ...441..436G}
have allowed to better understand the activity phenomenon beyond the Sun. A major contribution on
this subject was made by \citet{1998ASPC..154..153B}, who analysed the flux variability of
\ion{Ca}{ii} H\&K lines and found three different types of behaviour. 
They found cycles with periods between 2.5 and 25 yr often associated to stars with moderate activity.
Very active stars displayed fluctuations of activity rather than cycles, while inactive stars seem to be in a
state similar to the solar Maunder-minimum, the time period between the years 1645 and 1715, when the Sun
deviated from its usual 11-year activity cycle.  
  
Since pioneering surveys to date, several stellar cycles have been reported
\citep[e.g.][]{2007AJ....133..862H,2010ApJ...723L.213M,2010ApJ...722..343D,
2013ApJ...763L..26M,2014ApJ...781L...9B,2015ApJ...812...12E}. These works identified activity cycles in stars of
spectral types F to M (even stars with exoplanets), including stars with multiple cycles. 
The \ion{Ca}{ii} H\&K lines are commonly used as optical activity proxies
\citep[e.g. ][]{1990Natur.348..520B,1998ASPC..154..153B,2007AJ....133..862H}, as well as other
lines such as those of the Balmer series \citep{2013ApJ...764....3R, 2012AJ....143...93R} or the \ion{Mg}{ii} infrared
triplet \citep[see e.g.][]{2007A&A...466.1089B, 2011ApJ...736..114P}.
However, it is unclear if the activity cycles could be identified using other optical lines.
Such activity cycles have important implications in different fields.
For instance, studying a range of stars with physical characteristics similar to the Sun across
a range of age and other parameters is very useful to understand how typical the Sun is
\citep[e.g.][]{2007AJ....133..862H,2009AJ....138..312H},
and in particular the frequency of Earth-influencing behaviors (e.g. Maunder minimum).
Importantly, the discovery of activity cycles also helps to disentangle stellar and planetary signals
in radial-velocity surveys \citep[e.g.][]{2013ApJ...774..147R,2014A&A...567A..48C}.

We started a program aiming to detect activity cycles in solar-analog stars using the extensive database
of HARPS spectra. Our initial sample comprises close solar-analog stars taken from \citet{2015A&A...579A..52N},
who carefully selected stars with physical parameters very
similar\footnote{Strictly speaking, this is a sample of "solar-analog stars" rather than "solar-twins"
as was called by Nissen, given that (for instance) the age interval covered was 0.7 - 8.8 Gyr.}
to the Sun ($\pm$100 K in T$_\mathrm{eff}$, $\pm$0.15 dex in $\log g$, and $\pm$0.10 dex in [Fe/H])
and also requiring spectra with very high signal-to-noise S/N ($\ge$ 600).
The similarity between these stars and the Sun, together with the possibility to have spectra with
high S/N and an homogeneous parameter determination, encouraged us to initially select these objects.
However, we do not discard the future possible inclusion of more stars in order to extend our sample.

An inspection of the preliminary  data showed a notable variation in the chromospheric activity of
the G1.5V \citep{2006AJ....132..161G} star HD 45184 ($=$ HR 2318, V $=$ 6.39,
B$-$V $=$ 0.62). It is located at a distance of $\sim$ 22 pc \citep{2007A&A...474..653V} and hosts a
debris disk of 1-2 $M_\mathrm{\oplus}$ \citep{2009ApJ...705...89L}. 
The reported fundamental parameters are T$_\mathrm{eff} =$ 5871 $\pm$ 6 K , $\log g$ = 4.445 $\pm$ 0.012 dex, [Fe/H] $=$ 0.047 $\pm$ 0.006 dex and v$_\mathrm{turb} =$ 1.06 $\pm$ 0.017 km s$^{-1}$ \citep[][]{2015A&A...579A..52N}. He also derived a mass of 1.06 $\pm$ 0.01 M$_{\sun}$ (only internal error) and age of 2.7 $\pm$ 0.5 Gyr, showing that HD 45184 is $\sim$1.8 Gyr younger than the Sun. 
Then, with the aim to detect possible activity cycles, we collected all the available HARPS spectra
for this object. Multiple observations are required to properly identify both long and short-term
activity variations. For the case of HD 45184, we gathered a total of 291 spectra covering more than 10
years of observations, being then a very good target to search for a possible activity cycle.
HD 45184 has been also identified as a candidate to host a planetary mass companion of 0.04 M$_\mathrm{Jup}$
detected by radial velocity \citep{2011arXiv1109.2497M},
and is included in the Extrasolar Planets Encyclopaedia\footnote{http://exoplanet.eu/}.
These facts place HD 45184 as a very interesting object and encouraged us to perform the study
presented here.

This work is organized as follows. In Section §2 we describe
the observations and data reduction, while in Section §3 we describe our results and finally in Section §4 we outline our discussion.


\section{Observations and data reduction}

The spectra were obtained from the European Southern Observatory (ESO) archive\footnote{
https://www.eso.org/sci/facilities/lasilla/instruments/harps/tools/
archive.html},
under the ESO programs 072.C-0488(E) (PI: M. Mayor and S. Udry) and 183.C-0972(A)
(PI: S. Udry).
The observations were taken with the HARPS spectrograph attached to the La Silla 3.6m telescope
between 2003 to 2014 and have been automatically processed by the HARPS 
pipeline\footnote{\textsf{http://www.eso.org/sci/facilities/lasilla/instruments/harps/doc.html}}.
The spectra have a resolving power R$\sim$115\,000 and cover the spectral range  3782--6913~\AA.
After discarding a few low signal-to-noise observations we obtained a total of 291 spectra with a
mean signal-to-noise S/N $\sim$ 175 at 6490~\AA. 

Since this work was intended to search for low-level flux variations, the spectra were carefully
normalized and cleaned. Cosmic rays were identified and removed considering a threshold of 5 times
the noise level.
Continuum differences between the spectra were corrected by filtering low-frequency modulations.

\begin{figure}
   \centering
   \includegraphics[width=5cm,height=5cm, bb= 2 160 557 560, angle=-90 ]{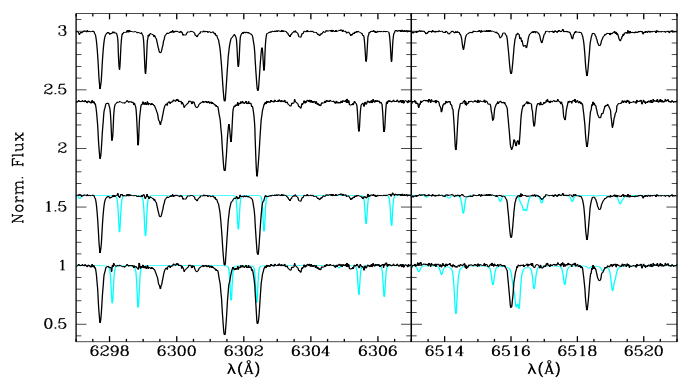} 
      \caption{Remotion of telluric lines. For two observations of different runs the original
      spectra are shown on the top. Below the clean spectra (black) are overplotted 
      with the telluric model spectra used in each case (light blue). 
      }
         \label{plot1}
   \end{figure}

Telluric features were removed dividing each observed spectrum $S_\mathrm{obs}$  by an empirical
telluric spectrum $T$ appropriately scaled:
$$
S_\mathrm{corr} = S_\mathrm{obs}\cdot T^{-\alpha},
$$
where $\alpha$ represents the relative optical thickness of the terrestrial atmosphere.
This coefficient $\alpha$ was chosen to reproduce the telluric line intensities, most of which
are due to water vapour in our spectral region. 
We repeat this procedure using a template of the O$_2$ $\gamma$-band around 6300 \AA. 
In this way we were able to correct satisfactorily observations taken with different atmospheric
water vapour content. Figure~\ref{plot1} shows an example of the removal of telluric lines in two
small spectral windows. Telluric lines in the left panel correspond to atmospheric O$_2$, while
in the right panel are due to H$_2$O. 

\section{Results}

\subsection{Chromospheric \ion{Ca}{ii} H\&K lines}
HARPS-S index was obtained following the classical method used to calculate the Mount Wilson S
index \citep{1978PASP...90..267V}. We integrated the flux in two windows centred at the cores
of the \ion{Ca}{ii} H\&K lines (3933.664~\AA~and 3968.470~\AA), weighting with triangular profiles
of 1.09 \AA~ FWHM.  The ratio of these fluxes to the mean continuum flux was computed by using 
two 20~\AA~wide passbands, centred at 3891\AA~and 4011\AA. 
Finally, we derived the Mount Wilson S index by using the calibration procedure explained in
\citet{2011arXiv1107.5325L}. 

\begin{figure}
\centering
\includegraphics[width=\hsize]{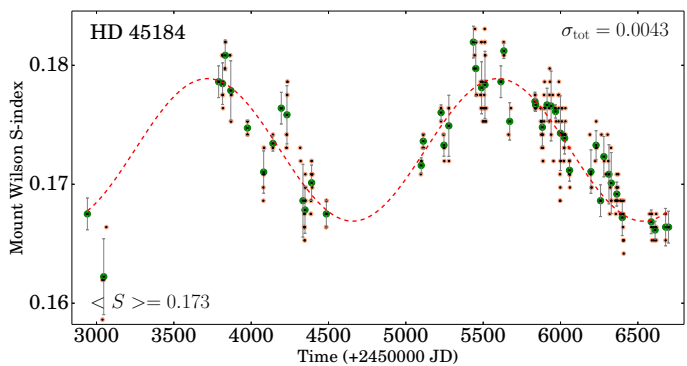}
\caption{Time series of Mount Wilson index measurements of HD 45184 from full HARPS data set  (small
orange circles) and seasonal means (big green circles). The dotted red line indicates the cycle
calculated in this work. The probable error of individual measurements is about 0.0043.}
\label{plot2}
\end{figure}

In Figure~\ref{plot2} we plot the time series of the original 291 HARPS measurements. 
The total uncertainty $\sigma_\mathrm{tot}$ of individual S determinations were derived by adding quadratically both the observational error $\sigma_\mathrm{obs}$ \citep[obtained as in][] {1996AJ....111..439H}, and the systematic error associated to the calibration between HARPS and Mt Wilson activity indexes \citep[$\sigma_\mathrm{sys}\sim 0.0043$, according to][]{2011arXiv1107.5325L}.
For clarity, error bars corresponding to $\sigma_\mathrm{tot}$ are not shown in Fig. 2, being almost identical for all points. 
In addition to the individual measurements, we include in the Fig.~\ref{plot2} the average values of observations associated to the same observing season. 
Error bars of the average values correspond to the standard deviation of the mean. 
For the case of bins with one measurement, we adopted the typical RMS dispersion
of other bins.
Both time series show a clear evidence of a chromospheric activity cycle. 
Applying the Lomb-Scargle periodogram \citep{1986ApJ...302..757H}  to the seasonal means, we derived a period P = 1878 $\pm 9$ days ($\sim$ 5.14 yr). We used seasonal means in this 
analysis in order to reduce the rotational scatter, as is usual \citep[e.g.][]{1995ApJ...438..269B,2011A&A...534A..30G}.
The False Alarm Probability (FAP) corresponding to the main peak is  3.1 $\times 10^{-07}$.

Along with the observational data, Fig.~\ref{plot2} shows the harmonic curve of period 1878 days
(red dotted line), obtained as the least-square fit to the seasonal means with a confidence level
of 99\% \citep[see][for details]{2009A&A...495..287B}. The semi-amplitude of the fitted curve is
$\Delta S=0.0062 \pm 0.0002$ and the epoch of maximum activity corresponds to JD 2,455,600. 

The mean Mount Wilson index of HD 45184 obtained in this work is $\left\langle S \right\rangle=0.173$
is undistinguishable from the values derived by 
\citet[][$\left\langle S \right\rangle=0.173$]{1996AJ....111..439H} and
\citet[][$\left\langle S \right\rangle=0.172$]{2006AJ....132..161G}. 
The mean activity level of HD 45184 is therefore  similar to that of the Sun
\citep[$\left\langle S \right\rangle_{\odot}=0.171$, ][]{2009AJ....138..312H}. 

\subsection{Balmer and metallic lines as activity indicators}
Several studies have suggested that the correlation between the chromospheric emission from the
\ion{Ca}{ii} H\&K and H$\alpha$ lines present in Sun \citep{2007ApJ...657.1137L} could be found in other stars 
\citep[e.g.][]
{2009A&A...501.1103M,2010A&A...520A..79M,2011MNRAS.414.2629M,2013A&A...558A.141S,2014A&A...566A..66G}. 
Even though this correlation has been reported as not always valid \citep{2007A&A...469..309C},
the activity in the Sun and other stars measured by \ion{Ca}{ii} H\&K emission could correlate with other chromospheric and even high photospheric lines
\citep[e.g.][]{1998ApJ...493..494H,2007ApJ...657.1137L}. 

To search for such correlations, 
we looked for possible variations in \ion{H}{i} and metallic lines in the spectra using different techniques. 
On the one hand we averaged the 291 spectra and built an RMS spectrum by calculating at each wavelength the RMS of residuals. Before these calculations, a mild filter in low and high-frequencies was applied to the residual spectra in order to remove continuum differences and reduce
shot noise. Since shot noise is an important contribution to the observed RMS, strong lines
present in general a RMS value lower  than the surroundings. For this reason, 
we divided the RMS spectrum by the theoretical noise-level spectrum calculated from photon statistics.  

As a second strategy, we applied the technique of \cite{2000A&A...353..707S}.
In this case, all the flux measurements at a given wavelength are used to produce a light curve that is fitted by least-squares with the first terms of a Fourier series.
In our study, we fixed the period and derived at each wavelength the amplitude of the 
flux variation assuming a sinusoidal function.

\begin{figure*}[]
\centering
\includegraphics[width=7cm,height=\hsize, bb=2 2 340 737, angle=-90]{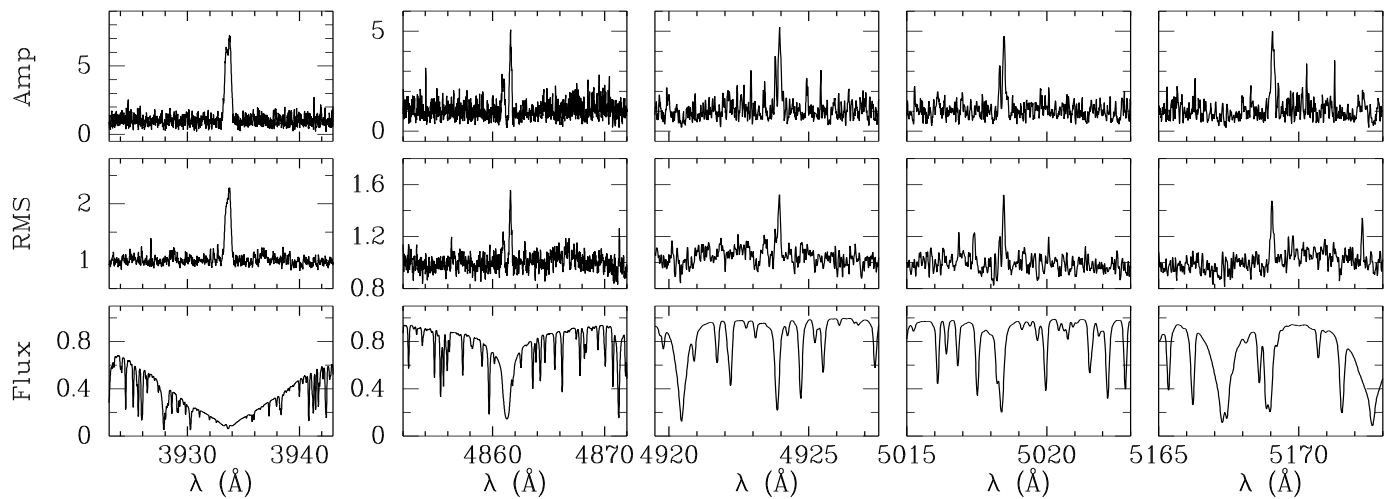}
\caption{Spectral variability. Light-variation amplitude spectrum (upper panels) and RMS 
spectrum (middle panels) are shown for small spectral windows around 5 variable
spectral lines: \ion{Ca}{ii} 3933, \ion{Fe}{ii} 4924, \ion{Fe}{ii} 5018, and \ion{Fe}{ii} 5169.
 Lower panels show the average spectrum in each region.} 
\label{fig:var}
\end{figure*}

From this analysis we detected, besides \ion{Ca}{ii} lines,
significant variations near the core of \ion{H}{i} lines and
in some strong iron lines, particularly  \ion{Fe}{ii} lines at 4924 \AA, 5018 {\AA} and 5169 \AA.
Figure~\ref{fig:var} shows the RMS spectrum and the light-variation amplitude spectrum 
for small spectral windows around selected lines.
The amplitude and RMS spectra have been normalized to the mean noise level in the region.
All mentioned lines (\ion{Ca}{ii} H\&K, H$\beta$, H$\alpha$, \ion{Fe}{ii} 4924 \AA, 5018 {\AA} and 5169 \AA)
present variations above 4$\sigma$, with  flux variations of about 0.6\% of the 
continuum level.
Additionally, a few other lines of ionized metals  exhibit variations at a 3$\sigma$ level,
which we consider here as marginal detections. These lines are the \ion{Ba}{ii} line at
6141.8 {\AA} and \ion{Ti}{ii} lines at 4395.0 \AA, 4443.8 \AA, 4468.5 \AA, 4501.3 {\AA}, and 4572.0 \AA.
In these cases, flux variations are about 0.4--0.5\% of the continuum.

\begin{figure}
\centering
\includegraphics[width=\hsize]{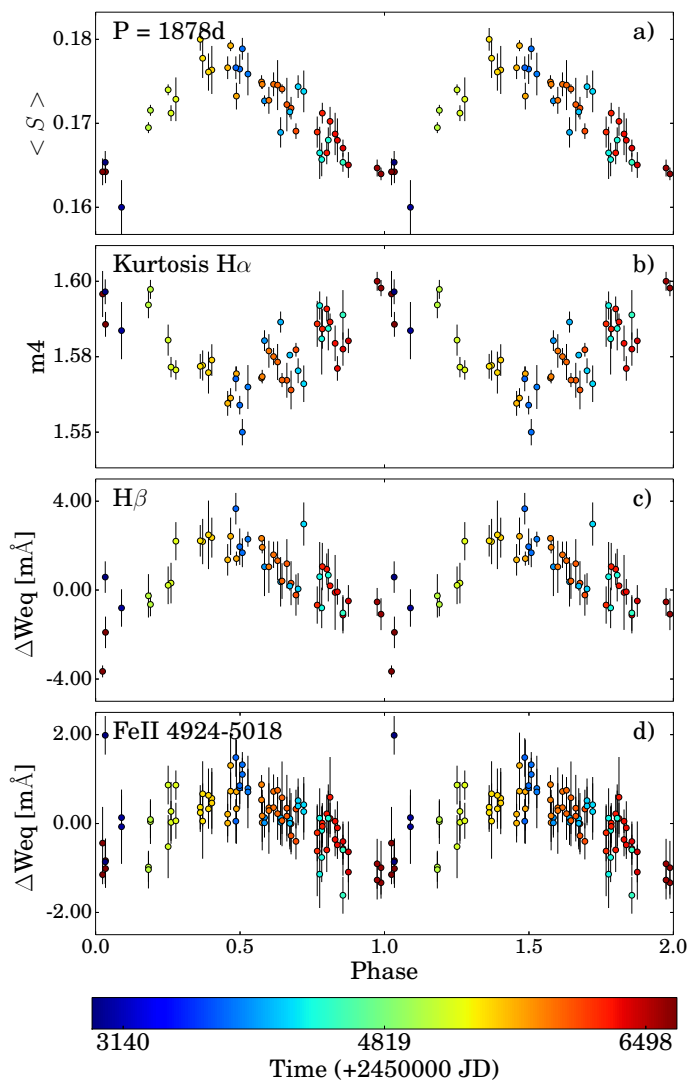}
\caption{a) Chromospheric activity measurements from \ion{Ca}{ii} H\&K lines. b) kurtosis  of the
H$\alpha$ line. c) and d) panels correspond to the equivalent width of Balmer H$\beta$, \ion{Fe}{ii}
4924 \AA~and \ion{Fe}{ii} 5018 \AA~lines. Phases have been calculated  with a period of 1878 days.}
\label{plot3}
\end{figure}

\begin{figure}
\centering
\includegraphics[width=11cm, height=6cm]{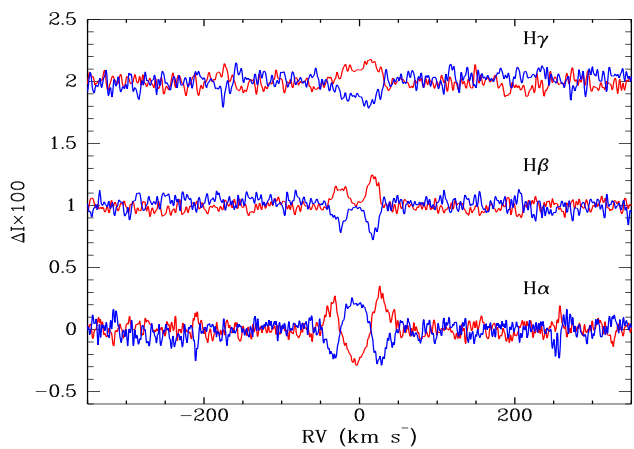}
\caption{Difference of Balmer line profiles for high (blue) and low (red) activity phases.}
\label{plot4}
\end{figure} 

\begin{figure}[]
\centering
\includegraphics[width=\hsize,height=6cm, bb=18 144 592 618]{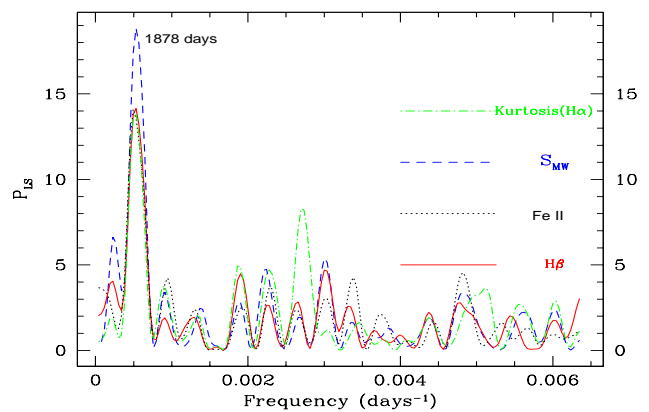}
\caption{Lomb-Scargle periodogram from the different line variations plotted in Fig. 4.}
\label{plot5}
\end{figure}

To study the temporal behaviour we derived equivalent width variations by
integrating the residual spectra around the central wavelength of the variable lines.
Figure~\ref{plot3} shows the Mount Wilson S index along with the equivalent width of \ion{H}{i} and
\ion{Fe}{ii} lines as a function of the activity cycle phase. 
There is a clear correlation between the intensity of the lines and the chromospheric activity,
for the lines H$\beta$, \ion{Fe}{ii} 4924 \AA, and \ion{Fe}{ii} 5018 \AA.
In the case of H$\alpha$, even though the spectral variability of the line core ($\pm$ 1~\AA~ around
the line center) is out of doubt, there are no noticeable net changes in the equivalent width.
This is evident in Fig.~\ref{plot4}, where we compare the average of 85 residual spectra close to the low
activity phase (blue) and the average of 57 spectra around the activity maximum (red). 
To quantify the spectral variations in H$\alpha$ we calculated the kurtosis of the line core, in an
spectral window of 4 \AA.
The resulting values show the same behaviour as other activity indicators as shown in the second panel
of Fig.~\ref{plot3}. Then, Balmer and \ion{Fe}{ii} lines variations reproduce the 5.14-yr activity
cycle detected with the Mount Wilson index, although with lower amplitude-to-error ratio. This is confirmed by the corresponding periodograms
(Fig.~\ref{plot5}) in which all the highest peaks lie between 1877 and 1879 days (the Mount Wilson indexes, the kurtosis of H$\alpha$, and the flux of H$\beta$ and \ion{Fe}{ii} lines).

\subsection{Rotation period}

As is often the case in other magnetic stars, the point spread in the chromospheric activity
curve is significantly larger than the estimated measurement errors, what can be attributed
to rotational modulations caused by passage of individual active regions 
\citep{1995ApJ...441..436G,2008ApJ...679.1531B,2010ApJ...723L.213M}. 
To detect such variations and to determine the rotation period, we have followed a strategy
similar to \citet{2010ApJ...723L.213M}. 
We subtract from the individual measurements of the S index the fitted harmonic curve (red dotted line in
Fig.~\ref{plot2}) and analyse the periodicity of the residuals by using using both
the Lomb-Scargle periodogram and the Phase Dispersion Minimization technique \citep{1978ApJ...224..953S}.

\begin{figure}
\centering
\includegraphics[width=\hsize,height=6cm, bb=18 140 592 618]{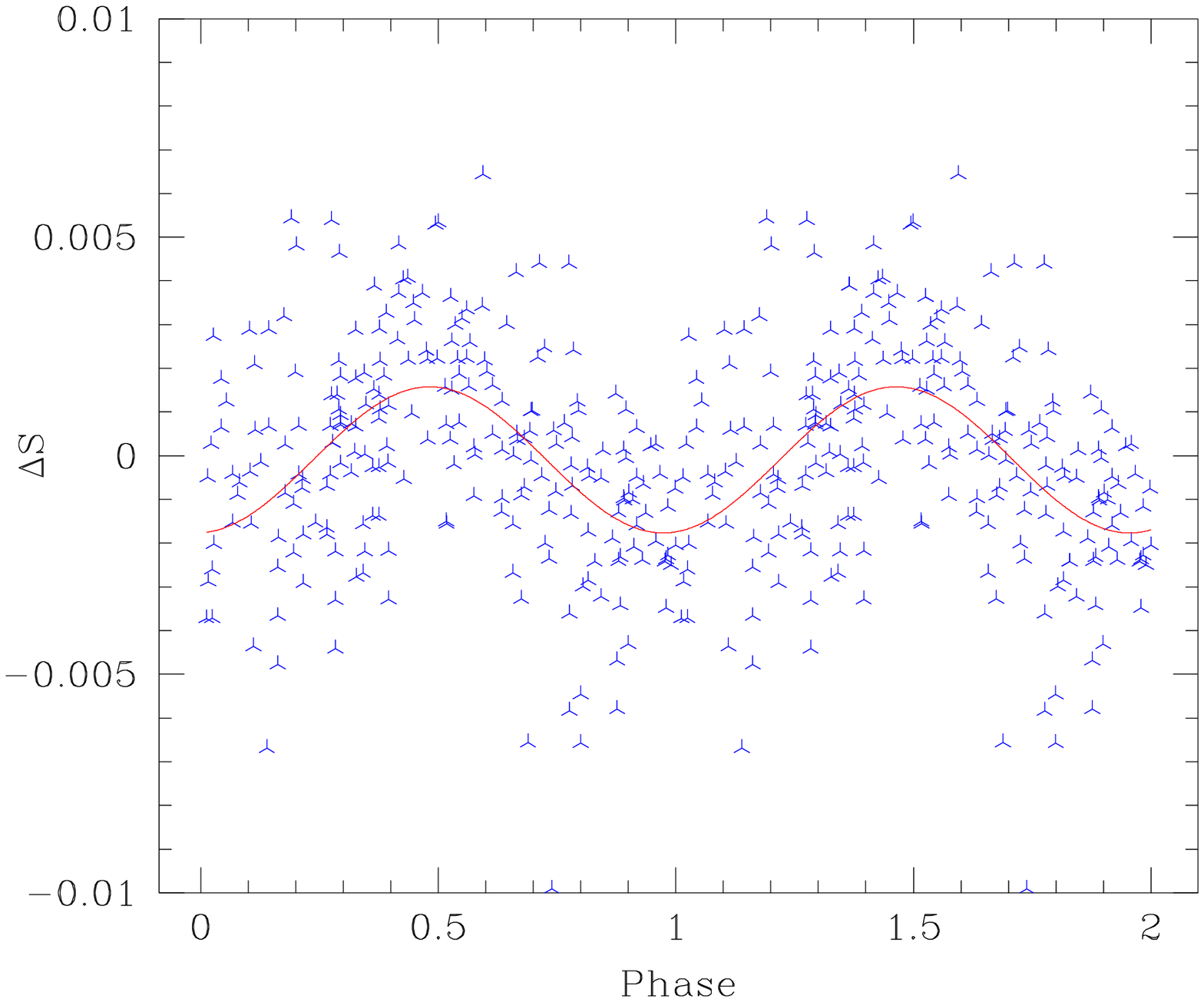}
\includegraphics[width=\hsize,height=6cm, bb=18 140 592 618]{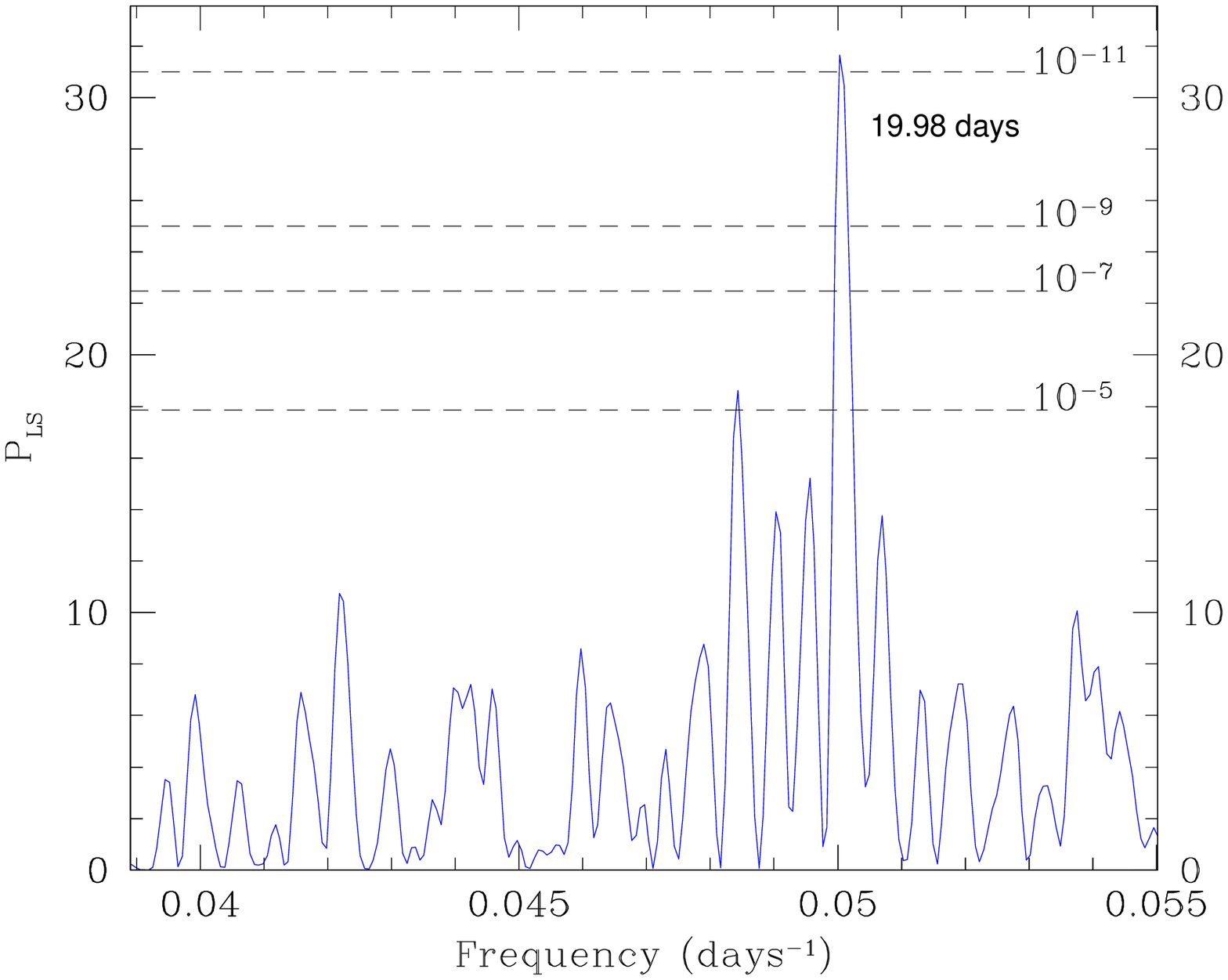}
\caption{Top panel: Mount Wilson index residuals phased with a period of $\sim$ 19.98 days. 
Lower panel: Lomb–Scargle periodogram from filtered (residuals) \ion{Ca}{ii} H\&K measurements. 
The highest peak suggests a rotation period 19.98 days.}
\label{plot6}
\end{figure}

As a result, we found a period of 19.98 $\pm$ 0.02 days with a very low FAP=1.04 $\times 10^{-11}$ for the S index.
This analysis was based in \ion{Ca}{ii} lines.
The signal-to-noise ratio of the variations found in other lines is too low to be
used for the rotational analysis. 
Figure~\ref{plot6} shows the periodogram and the S-index residuals as a function of phase. Our result is in agreement with the period estimated  by \citet{2004ApJS..152..261W} (21 days) using the empirical relation between rotation and chromospheric activity calibrated by \citet{1984ApJ...279..763N}.

\section{Discussion} 

We studied the long-term activity of the solar-analog star HD 45184 by using the
\ion{Ca}{ii} H\&K and Balmer lines as stellar activity proxies.
We analysed the \ion{Ca}{ii} H\&K line-core fluxes in 291 HARPS spectra taken between 2003 and 2014. 
In addition, we detected spectral variability in the first 3 lines of the Balmer series
and in some strong lines of ionized metals, particularly \ion{Fe}{ii} lines at 4924 \AA, 
 5018 \AA, and 5169 \AA.
We detected a long-term activity cycle of 5.14-yr from the \ion{Ca}{ii} H\&K lines, which is clearly
replicated by line profile variations in Balmer and strong \ion{Fe}{ii} lines (see e.g. Fig.~\ref{plot3}).
Equivalent width variations of Balmer (except for H$\alpha$) and \ion{Fe}{ii} lines 
present a positive correlation with the emission in the \ion{Ca}{ii} H\&K lines:
the more intense the emission of the \ion{Ca}{ii} H\&K lines, the stronger the \ion{H}{i} and \ion{Fe}{ii}
lines. In the case of H$\alpha$, the equivalent width does not suffer significant changes,
although the line profile shape varies in phase with the activity cycle.   
 
It is believed that the surface of older and less active stars, including our Sun, are dominated by
bright regions called faculae, whereas young and active stars are spot dominated \citep{1998ApJS..118..239R,2002A&A...394..505B}.
\citet{1966sst..book.....M} showed a number of Fe lines sensitive to the presence of Sun spots.
Based on this scenario and considering the similarities between the Sun and HD~45184, an \ion{Fe}{ii}
variation following the Ca H\&K activity cycle could be possibly attributed to the lower age
of HD 45184. This could be verified using future observations of other young solar-analog stars.

Solar spectral variation of several Fraunhofer lines along the 11-yr activity cycle were first
revealed by \citet{1982ApJ...252..375L}. Next, taking into account that these lines are formed at
different depths in the atmosphere of the Sun, \citet{2007ApJ...657.1137L} used some
photospheric and chromospheric lines to study the long-term variations of the temperature.
Monitoring high photospheric lines from 1980 to 2006, they found that Mn 5394 \AA \ is the only
photospheric line which shows a clear modulation with the chromospheric 11-yr \ion{Ca}{ii} K cycle
of the Sun. 
Surprisingly, Fe lines did not show the expected modulation with \ion{Ca}{ii} H\&K lines, and then they suggest that these lines
are probably following the 22-yr Hale cycle. Interestingly, in HD~45184 we found that \ion{Fe}{ii} lines show a clear modulation
with the 5.14-yr \ion{Ca}{ii} H\&K activity cycle. 
Following \citet{1966sst..book.....M}, the variations of these high photospheric \ion{Fe}{ii}
lines could point toward the possible presence of dark spots in the surface of HD 45184. At the same time, and similar to the solar case, the faculae are presumably present on the stellar surface, as suggested by the \ion{Ca}{ii} H\&K emission lines \citep[e.g.][and references therein]{2008LRSP....5....2H}.

Using the \ion{Ca}{ii} H\&K lines we found for short-term modulations caused by stellar rotation
and determined a rotation period of 19.98 days for HD 45184. This value is in agreement with the
rotational period vs. chromospheric activity calibration of \citet{2004ApJS..152..261W}.
On the other hand, being HD 45184 more active than the Sun, the rotation period calculated by us is
expected to be shorter than the rotation period of Sun, because of the widely believed close
connection between rotation and activity \citep{1984ApJ...279..763N}.

\begin{figure}[]
\centering
\includegraphics[width=7cm,height=9cm, bb=2 2 557 737, angle=-90]{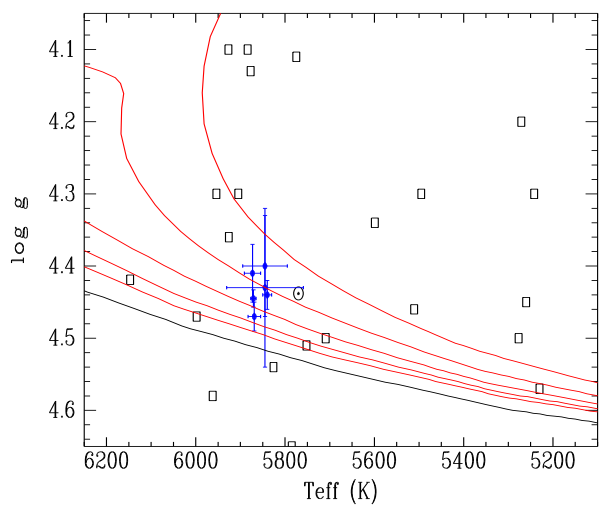}
\caption{Position of G-type stars with activity cycles in the T$_\mathrm{eff}-\log g$ diagram (black empty squares). 
Blue filled circles with error bars correspond to the parameters of HD~45184 derived by different authors. Red lines are isochrones for solar abundances are shown for $\log$ age = 9.0, 9.2, 9.4, 9.6, and 9.8; black line is the ZAMS. The position of the Sun is also included in the figure.}
\label{fig:tlogg}
\end{figure}

Despite solar-twin and solar-analog stars are ideal laboratories to carry out comparative 
studies of stellar cycles and the solar cycle, not many of them have been 
considered as targets of activity studies.
An interesting aspect of HD\,45184 is that, among  stars with detected activity cycles,
it is one of which have stellar parameters closest to the solar values.
In Fig.~\ref{fig:tlogg} we show the position in the T$_\mathrm{eff}-\log g$ diagram of stars with known activity cycles.
We have included  G-type stars studied by \citet{2015ApJ...802...67C} and
the Mount Wilson project stars selected by \citet{2013A&A...554A..50S} as having  cyclic 
variations. Additionally, we added individual objects with activity periods 
reported by  \citet{2010ApJ...723L.213M}, \citet{2010ApJ...722..343D}, and \citet{2015ApJ...812...12E}.
We included in the Figure the parameters of HD~45184 determined by different literature works \citep{2008A&A...487..373S,2014A&A...562A..71B,2015A&A...577A...9B,2015A&A...579A..20M,2015A&A...579A..52N,2016A&A...585A.152S}, using blue filled circles with error bars. For comparison, theoretical isochrones of \citet{2012A&A...541A..41M} for solar abundances are plotted.
HD~45184 has almost the same surface gravity as the Sun within the uncertainties and it is slightly
hotter. 
In particular, \citet{2014A&A...562A..71B} estimate an age of 4.4$\pm$2.2 Gyr, similar to the age of the Sun. However, the parameters reported by \citet[][]{2008A&A...487..373S} suggest that HD 45184 is about 2 Gyr  younger than the Sun, in agreement with \citet{2015A&A...579A..52N}. Without going in further detail, we can say that HD~45184 is $\sim$ 100 K hotter than the
Sun, and is one of the closest solar-analog stars with a known activity cycle.

It would be interesting to study if the \ion{Fe}{ii} lines could be also useful, for instance, to detect a possible activity cycle in those planet-search radial-velocity surveys that do not include the classical \ion{Ca}{ii} H\&K lines in their spectra. This is the case, for example, of the Anglo-Australian Planet Search survey \citep[e.g.][]{1996PASP..108..500B,2001ApJ...555..410B,2011ApJ...727..103T}. We caution, however, that more observations are needed to verify the usefulness of the Fe and possibly other metallic lines in solar-analog stars.

\begin{acknowledgements} 
We thank the anonymous referee for constructive comments that improved the paper. 
\end{acknowledgements} 

\bibliographystyle{aa}
\bibliography{references}

\begin{thebibliography}{59}
\expandafter\ifx\csname natexlab\endcsname\relax\def\natexlab#1{#1}\fi

\bibitem[{{Baliunas} \& {Jastrow}(1990)}]{1990Natur.348..520B}
{Baliunas}, S. \& {Jastrow}, R. 1990, \nat, 348, 520

\bibitem[{{Baliunas} {et~al.}(1995){Baliunas}, , \& {et
  al.}}]{1995ApJ...438..269B}
{Baliunas}, S.~L., , \& {et al.} 1995, Astrophys. J., 438, 269

\bibitem[{{Baliunas} {et~al.}(1998){Baliunas}, {Donahue}, {Soon}, \&
  {Henry}}]{1998ASPC..154..153B}
{Baliunas}, S.~L., {Donahue}, R.~A., {Soon}, W., \& {Henry}, G.~W. 1998, in ASP
  Conf. Ser. 154: Cool Stars, Stellar Systems, and the Sun, ed. R.~A. {Donahue}
  \& J.~A. {Bookbinder}, 153--+

\bibitem[{{Battistini} \& {Bensby}(2015)}]{2015A&A...577A...9B}
{Battistini}, C. \& {Bensby}, T. 2015, \aap, 577, A9

\bibitem[{{Bensby} {et~al.}(2014){Bensby}, {Feltzing}, \&
  {Oey}}]{2014A&A...562A..71B}
{Bensby}, T., {Feltzing}, S., \& {Oey}, M.~S. 2014, \aap, 562, A71

\bibitem[{{Berdyugina} {et~al.}(2002){Berdyugina}, {Pelt}, \&
  {Tuominen}}]{2002A&A...394..505B}
{Berdyugina}, S.~V., {Pelt}, J., \& {Tuominen}, I. 2002, \aap, 394, 505

\bibitem[{{Brown} {et~al.}(2008){Brown}, {Gray}, \&
  {Baliunas}}]{2008ApJ...679.1531B}
{Brown}, K.~I.~T., {Gray}, D.~F., \& {Baliunas}, S.~L. 2008, \apj, 679, 1531

\bibitem[{{Buccino} \& {Mauas}(2009)}]{2009A&A...495..287B}
{Buccino}, A.~P. \& {Mauas}, P.~J.~D. 2009, \aap, 495, 287

\bibitem[{{Buccino} {et~al.}(2014){Buccino}, {Petrucci}, {Jofr{\'e}}, \&
  {Mauas}}]{2014ApJ...781L...9B}
{Buccino}, A.~P., {Petrucci}, R., {Jofr{\'e}}, E., \& {Mauas}, P.~J.~D. 2014,
  \apjl, 781, L9

\bibitem[{{Bus{\`a}} {et~al.}(2007){Bus{\`a}}, {Aznar Cuadrado}, {Terranegra},
  {Andretta}, \& {Gomez}}]{2007A&A...466.1089B}
{Bus{\`a}}, I., {Aznar Cuadrado}, R., {Terranegra}, L., {Andretta}, V., \&
  {Gomez}, M.~T. 2007, \aap, 466, 1089

\bibitem[{{Butler} {et~al.}(1996){Butler}, {Marcy}, {Williams}, {McCarthy},
  {Dosanjh}, \& {Vogt}}]{1996PASP..108..500B}
{Butler}, R.~P., {Marcy}, G.~W., {Williams}, E., {et~al.} 1996, \pasp, 108, 500

\bibitem[{{Butler} {et~al.}(2001){Butler}, {Tinney}, {Marcy}, {Jones}, {Penny},
  \& {Apps}}]{2001ApJ...555..410B}
{Butler}, R.~P., {Tinney}, C.~G., {Marcy}, G.~W., {et~al.} 2001, \apj, 555, 410

\bibitem[{{Carolo} {et~al.}(2014){Carolo}, {Desidera}, {Gratton}, {Martinez
  Fiorenzano}, {Marzari}, {Endl}, {Mesa}, {Barbieri}, {Cecconi}, {Claudi},
  {Cosentino}, \& {Scuderi}}]{2014A&A...567A..48C}
{Carolo}, E., {Desidera}, S., {Gratton}, R., {et~al.} 2014, \aap, 567, A48

\bibitem[{{Choi} {et~al.}(2015){Choi}, {Lee}, {Oh}, {Kim}, {Kim}, \&
  {Yi}}]{2015ApJ...802...67C}
{Choi}, H., {Lee}, J., {Oh}, S., {et~al.} 2015, \apj, 802, 67

\bibitem[{{Cincunegui} {et~al.}(2007){Cincunegui}, {D{\'{\i}}az}, \&
  {Mauas}}]{2007A&A...469..309C}
{Cincunegui}, C., {D{\'{\i}}az}, R., \& {Mauas}, P. 2007, \aap, 469, 309

\bibitem[{{DeWarf} {et~al.}(2010){DeWarf}, {Datin}, \&
  {Guinan}}]{2010ApJ...722..343D}
{DeWarf}, L.~E., {Datin}, K.~M., \& {Guinan}, E.~F. 2010, \apj, 722, 343

\bibitem[{{Duncan} {et~al.}(1991){Duncan}, {Vaughan}, {Wilson}, {Preston},
  {Frazer}, {Lanning}, {Misch}, {Mueller}, {Soyumer}, {Woodard}, {Baliunas},
  {Noyes}, {Hartmann}, {Porter}, {Zwaan}, {Middelkoop}, {Rutten}, \&
  {Mihalas}}]{1991ApJS...76..383D}
{Duncan}, D.~K., {Vaughan}, A.~H., {Wilson}, O.~C., {et~al.} 1991, \apjs, 76,
  383

\bibitem[{{Egeland} {et~al.}(2015){Egeland}, {Metcalfe}, {Hall}, \&
  {Henry}}]{2015ApJ...812...12E}
{Egeland}, R., {Metcalfe}, T.~S., {Hall}, J.~C., \& {Henry}, G.~W. 2015, \apj,
  812, 12

\bibitem[{{Gomes da Silva} {et~al.}(2014){Gomes da Silva}, {Santos}, {Boisse},
  {Dumusque}, \& {Lovis}}]{2014A&A...566A..66G}
{Gomes da Silva}, J., {Santos}, N.~C., {Boisse}, I., {Dumusque}, X., \&
  {Lovis}, C. 2014, \aap, 566, A66

\bibitem[{{Gomes da Silva} {et~al.}(2011){Gomes da Silva}, {Santos}, {Bonfils},
  {Delfosse}, {Forveille}, \& {Udry}}]{2011A&A...534A..30G}
{Gomes da Silva}, J., {Santos}, N.~C., {Bonfils}, X., {et~al.} 2011, \aap, 534,
  A30

\bibitem[{{Gray} \& {Baliunas}(1995)}]{1995ApJ...441..436G}
{Gray}, D.~F. \& {Baliunas}, S.~L. 1995, \apj, 441, 436

\bibitem[{{Gray} {et~al.}(2006){Gray}, {Corbally}, {Garrison}, {McFadden},
  {Bubar}, {McGahee}, {O'Donoghue}, \& {Knox}}]{2006AJ....132..161G}
{Gray}, R.~O., {Corbally}, C.~J., {Garrison}, R.~F., {et~al.} 2006, \aj, 132,
  161

\bibitem[{{Hall}(2008)}]{2008LRSP....5....2H}
{Hall}, J.~C. 2008, Living Reviews in Solar Physics, 5

\bibitem[{{Hall} {et~al.}(2009){Hall}, {Henry}, \&
  {Lockwood}}]{2009AJ....138..312H}
{Hall}, J.~C., {Henry}, G.~W., \& {Lockwood}, G.~W. 2009, \aj, 138, 312

\bibitem[{{Hall} \& {Lockwood}(1998)}]{1998ApJ...493..494H}
{Hall}, J.~C. \& {Lockwood}, G.~W. 1998, \apj, 493, 494

\bibitem[{{Hall} {et~al.}(2007){Hall}, {Lockwood}, \&
  {Skiff}}]{2007AJ....133..862H}
{Hall}, J.~C., {Lockwood}, G.~W., \& {Skiff}, B.~A. 2007, \aj, 133, 862

\bibitem[{{Henry} {et~al.}(1996){Henry}, {Soderblom}, {Donahue}, \&
  {Baliunas}}]{1996AJ....111..439H}
{Henry}, T.~J., {Soderblom}, D.~R., {Donahue}, R.~A., \& {Baliunas}, S.~L.
  1996, \aj, 111, 439

\bibitem[{{Horne} \& {Baliunas}(1986)}]{1986ApJ...302..757H}
{Horne}, J.~H. \& {Baliunas}, S.~L. 1986, \apj, 302, 757

\bibitem[{{Lawler} {et~al.}(2009){Lawler}, {Beichman}, {Bryden}, {Ciardi},
  {Tanner}, {Su}, {Stapelfeldt}, {Lisse}, \& {Harker}}]{2009ApJ...705...89L}
{Lawler}, S.~M., {Beichman}, C.~A., {Bryden}, G., {et~al.} 2009, \apj, 705, 89

\bibitem[{{Livingston} \& {Holweger}(1982)}]{1982ApJ...252..375L}
{Livingston}, W. \& {Holweger}, H. 1982, \apj, 252, 375

\bibitem[{{Livingston} {et~al.}(2007){Livingston}, {Wallace}, {White}, \&
  {Giampapa}}]{2007ApJ...657.1137L}
{Livingston}, W., {Wallace}, L., {White}, O.~R., \& {Giampapa}, M.~S. 2007,
  \apj, 657, 1137

\bibitem[{{Lovis} {et~al.}(2011){Lovis}, {Dumusque}, {Santos}, {Bouchy},
  {Mayor}, {Pepe}, {Queloz}, {S{\'e}gransan}, \& {Udry}}]{2011arXiv1107.5325L}
{Lovis}, C., {Dumusque}, X., {Santos}, N.~C., {et~al.} 2011, \arxiv
  [\eprint[arXiv]{1107.5325}]

\bibitem[{{Maldonado} {et~al.}(2015){Maldonado}, {Eiroa}, {Villaver},
  {Montesinos}, \& {Mora}}]{2015A&A...579A..20M}
{Maldonado}, J., {Eiroa}, C., {Villaver}, E., {Montesinos}, B., \& {Mora}, A.
  2015, \aap, 579, A20

\bibitem[{{Mart{\'{\i}}nez-Arn{\'a}iz}
  {et~al.}(2011){Mart{\'{\i}}nez-Arn{\'a}iz}, {L{\'o}pez-Santiago},
  {Crespo-Chac{\'o}n}, \& {Montes}}]{2011MNRAS.414.2629M}
{Mart{\'{\i}}nez-Arn{\'a}iz}, R., {L{\'o}pez-Santiago}, J.,
  {Crespo-Chac{\'o}n}, I., \& {Montes}, D. 2011, \mnras, 414, 2629

\bibitem[{{Mart{\'{\i}}nez-Arn{\'a}iz}
  {et~al.}(2010){Mart{\'{\i}}nez-Arn{\'a}iz}, {Maldonado}, {Montes}, {Eiroa},
  \& {Montesinos}}]{2010A&A...520A..79M}
{Mart{\'{\i}}nez-Arn{\'a}iz}, R., {Maldonado}, J., {Montes}, D., {Eiroa}, C.,
  \& {Montesinos}, B. 2010, \aap, 520, A79

\bibitem[{{Mayor} {et~al.}(2011){Mayor}, {Marmier}, {Lovis}, {Udry},
  {S{\'e}gransan}, {Pepe}, {Benz}, {Bertaux}, {Bouchy}, {Dumusque}, {Lo Curto},
  {Mordasini}, {Queloz}, \& {Santos}}]{2011arXiv1109.2497M}
{Mayor}, M., {Marmier}, M., {Lovis}, C., {et~al.} 2011, ArXiv e-prints
  [\eprint[ArXiv Astrophysics e-prints]{1109.2497}]

\bibitem[{{Metcalfe} {et~al.}(2010){Metcalfe}, {Basu}, {Henry}, {Soderblom},
  {Judge}, {Kn{\"o}lker}, {Mathur}, \& {Rempel}}]{2010ApJ...723L.213M}
{Metcalfe}, T., {Basu}, S., {Henry}, T.~J., {et~al.} 2010, \apjl, 723, L213

\bibitem[{{Metcalfe} {et~al.}(2013){Metcalfe}, {Buccino}, {Brown}, {Mathur},
  {Soderblom}, {Henry}, {Mauas}, {Petrucci}, {Hall}, \&
  {Basu}}]{2013ApJ...763L..26M}
{Metcalfe}, T., {Buccino}, A., {Brown}, B., {et~al.} 2013, \apjl, 763, L26

\bibitem[{{Meunier} \& {Delfosse}(2009)}]{2009A&A...501.1103M}
{Meunier}, N. \& {Delfosse}, X. 2009, \aap, 501, 1103

\bibitem[{{Moore} {et~al.}(1966){Moore}, {Minnaert}, \&
  {Houtgast}}]{1966sst..book.....M}
{Moore}, C.~E., {Minnaert}, M.~G.~J., \& {Houtgast}, J. 1966, {The solar
  spectrum 2935 A to 8770 A}

\bibitem[{{Mowlavi} {et~al.}(2012){Mowlavi}, {Eggenberger}, {Meynet},
  {Ekstr{\"o}m}, {Georgy}, {Maeder}, {Charbonnel}, \&
  {Eyer}}]{2012A&A...541A..41M}
{Mowlavi}, N., {Eggenberger}, P., {Meynet}, G., {et~al.} 2012, \aap, 541, A41

\bibitem[{{Nissen}(2015)}]{2015A&A...579A..52N}
{Nissen}, P.~E. 2015, \aap, 579, A52

\bibitem[{{Noyes} {et~al.}(1984){Noyes}, {Hartmann}, {Baliunas}, {Duncan}, \&
  {Vaughan}}]{1984ApJ...279..763N}
{Noyes}, R.~W., {Hartmann}, L.~W., {Baliunas}, S.~L., {Duncan}, D.~K., \&
  {Vaughan}, A.~H. 1984, \apj, 279, 763

\bibitem[{{Pietarila} \& {Livingston}(2011)}]{2011ApJ...736..114P}
{Pietarila}, A. \& {Livingston}, W. 2011, \apj, 736, 114

\bibitem[{{Radick} {et~al.}(1998){Radick}, {Lockwood}, {Skiff}, \&
  {Baliunas}}]{1998ApJS..118..239R}
{Radick}, R.~R., {Lockwood}, G.~W., {Skiff}, B.~A., \& {Baliunas}, S.~L. 1998,
  \apjs, 118, 239

\bibitem[{{Reiners} {et~al.}(2012){Reiners}, {Joshi}, \&
  {Goldman}}]{2012AJ....143...93R}
{Reiners}, A., {Joshi}, N., \& {Goldman}, B. 2012, \aj, 143, 93

\bibitem[{{Robertson} {et~al.}(2013{\natexlab{a}}){Robertson}, {Endl},
  {Cochran}, \& {Dodson-Robinson}}]{2013ApJ...764....3R}
{Robertson}, P., {Endl}, M., {Cochran}, W.~D., \& {Dodson-Robinson}, S.~E.
  2013{\natexlab{a}}, \apj, 764, 3

\bibitem[{{Robertson} {et~al.}(2013{\natexlab{b}}){Robertson}, {Endl},
  {Cochran}, {MacQueen}, \& {Boss}}]{2013ApJ...774..147R}
{Robertson}, P., {Endl}, M., {Cochran}, W.~D., {MacQueen}, P.~J., \& {Boss},
  A.~P. 2013{\natexlab{b}}, \apj, 774, 147

\bibitem[{{Schr{\"o}der} {et~al.}(2013){Schr{\"o}der}, {Mittag}, {Hempelmann},
  {Gonz{\'a}lez-P{\'e}rez}, \& {Schmitt}}]{2013A&A...554A..50S}
{Schr{\"o}der}, K.-P., {Mittag}, M., {Hempelmann}, A.,
  {Gonz{\'a}lez-P{\'e}rez}, J.~N., \& {Schmitt}, J.~H.~M.~M. 2013, \aap, 554,
  A50

\bibitem[{{Sokolov}(2000)}]{2000A&A...353..707S}
{Sokolov}, N.~A. 2000, \aap, 353, 707

\bibitem[{{Sousa} {et~al.}(2008){Sousa}, {Santos}, {Mayor}, {Udry},
  {Casagrande}, {Israelian}, {Pepe}, {Queloz}, \&
  {Monteiro}}]{2008A&A...487..373S}
{Sousa}, S.~G., {Santos}, N.~C., {Mayor}, M., {et~al.} 2008, \aap, 487, 373

\bibitem[{{Spina} {et~al.}(2016){Spina}, {Mel{\'e}ndez}, \&
  {Ram{\'{\i}}rez}}]{2016A&A...585A.152S}
{Spina}, L., {Mel{\'e}ndez}, J., \& {Ram{\'{\i}}rez}, I. 2016, \aap, 585, A152

\bibitem[{{Stellingwerf}(1978)}]{1978ApJ...224..953S}
{Stellingwerf}, R.~F. 1978, \apj, 224, 953

\bibitem[{{Stelzer} {et~al.}(2013){Stelzer}, {Frasca}, {Alcal{\'a}}, {Manara},
  {Biazzo}, {Covino}, {Rigliaco}, {Testi}, {Covino}, \&
  {D'Elia}}]{2013A&A...558A.141S}
{Stelzer}, B., {Frasca}, A., {Alcal{\'a}}, J.~M., {et~al.} 2013, \aap, 558,
  A141

\bibitem[{{Tinney} {et~al.}(2011){Tinney}, {Butler}, {Jones}, {Wittenmyer},
  {O'Toole}, {Bailey}, \& {Carter}}]{2011ApJ...727..103T}
{Tinney}, C.~G., {Butler}, R.~P., {Jones}, H.~R.~A., {et~al.} 2011, \apj, 727,
  103

\bibitem[{{van Leeuwen}(2007)}]{2007A&A...474..653V}
{van Leeuwen}, F. 2007, \aap, 474, 653

\bibitem[{{Vaughan} {et~al.}(1978){Vaughan}, {Preston}, \&
  {Wilson}}]{1978PASP...90..267V}
{Vaughan}, A.~H., {Preston}, G.~W., \& {Wilson}, O.~C. 1978, \mnras, 90, 267

\bibitem[{{Wilson}(1978)}]{1978ApJ...226..379W}
{Wilson}, O.~C. 1978, \apj, 226, 379

\bibitem[{{Wright} {et~al.}(2004){Wright}, {Marcy}, {Butler}, \&
  {Vogt}}]{2004ApJS..152..261W}
{Wright}, J.~T., {Marcy}, G.~W., {Butler}, R.~P., \& {Vogt}, S.~S. 2004, \apjs,
  152, 261

\end{thebibliography}

\end{document}